\documentclass[aps,prd,superscriptaddress,nofootinbib,amsmath,amsfonts,preprintnumbers,groupedaddress,showpacs,10pt,english]{revtex4-1}
\usepackage{amsmath}
\usepackage{amssymb}
\usepackage{babel}
\usepackage{wrapfig}
\usepackage{cancel}

\usepackage{relsize,exscale}
\makeatletter

\usepackage{array,multirow,graphicx}
\usepackage{dcolumn}
\usepackage{newlfont}
\usepackage{bm}
\usepackage[colorlinks,citecolor=blue,urlcolor=blue,linkcolor=blue]{hyperref}
\usepackage[figtopcap]{subfigure}
\usepackage{color}

\newcommand{\arctanh}{\mathrm{arctanh}}

\graphicspath{{./Figs/}}

\begin{document}

\date{\today}

\title{ Isotropic  stellar model in mimetic theory}

\author{G.G.L. Nashed}
\email{nashed@bue.edu.eg}
\affiliation {Centre for Theoretical Physics, The British University in Egypt, P.O. Box
43, El Sherouk City, Cairo 11837, Egypt}

\begin{abstract}
We investigate  how to derive an isotropic stellar model in the framework
of mimetic gravitational theory. Recently, this theory has gained  big interest due to its difference from Einstein's general relativity (GR), especially in the domain non-vacuum solutions.  In this regard, we apply the field equation of mimetic gravitational theory to a spherically symmetric ansatz and obtain an over determined system of non-linear differential equations in which differential equations are less than the unknown functions.  To overcome the over determined system we suppose a specific form of the temporal component of the metric potential, $g_{tt}$,  and assume the vanishing of the anisotropic condition to derive the form of the spatial component of the metric, $g_{rr}$. In this regard, we discuss the possibility to derive a stellar isotropic model that is in agreement with observed pulsars.  To examine the stability of the isotropic model we use the Tolman-Oppenheimer-Volkoff equation and the adiabatic index.  Furthermore, we assess the model's validity by evaluating its compatibility with a broad range of observed pulsar masses and radii. We demonstrate that the model provides a good fit to these observations.
\keywords{Mimetic theory;  Isotropic model; Tolman-Oppenheimer-Volkoff equation, adiabatic index.}
\end{abstract}

\maketitle
\section{Introduction}\label{S1}

{ The theory of  General Relativity (GR) was constructed by Einstein in (1915) and is considered one of the basic theories of modern physics as well as the quantum field theory \cite{carroll2004spacetime}. Up to date, GR has approved many successful tests in experimental as well as observational like gravitational time dilation, bending of light, the precession of the Mercury orbit, gravitational lensing, etc \cite{Will:2005va}, and the  discovery of the gravitational waves  \cite{LIGOScientific:2016aoc}. In spite the huge progress of GR, it endures investigating the issues of cosmological observations like the flat galaxy's rotation curves (dark matter), the black holes singularities as well as the accelerated expansion era of the universe (dark energy). Thus, new components of matter-energy or modified theories of gravity should be proposed to investigate the observed events.}

 Mimetic gravitational theory is a scalar-tensor one where the conformal mode can be isolated through a scalar field \cite{Chamseddine:2013kea}. On the other hand, we can think of the setup of the mimetic as a special class of
general conformal or disformal transformation where the transformation between the new and old
metrics is degenerate. Using the non-invertible conformal or disformal transformation one can prove that the number of
degrees of freedom can be increased so that the longitudinal mode becomes dynamical
\cite{Deruelle:2014zza,Domenech:2015tca,Firouzjahi:2018xob,Shen:2019nyp}. The conformal transformation which relates the  auxiliary metric ${\bar g}_{\alpha \beta}$  to the physical metric $g_{\alpha \beta}$
and the scalar field is defined as:
\begin{align}
\label{trans1}
g_{\alpha\beta}=\pm \left(\bar{g}^{\mu \nu} \partial_\mu \zeta\partial_\nu \zeta \right) \bar{g}_{\alpha\beta}\,.
\end{align}
We stress that the physical metric $g_{\alpha\beta}$ is invariant using the conformal transformation of the auxiliary
metric $\bar{g}^{\alpha \beta}$. This invariance fixes in a  unique way the form of the conformal factor w.r.t. the auxiliary metric  $\bar{g}^{\mu \nu} $  and the scalar field $\zeta$  however such transformation   cannot fix the sign.  Equation (\ref{trans1}) yields that the  following condition:
\begin{align}
\label{trans2}
g^{\alpha \beta}\partial_\alpha \zeta \partial_\beta \zeta= \pm 1\,.
\end{align}
Equation (\ref{trans2}) shows that $\partial_\beta \zeta$ is a timelike for the $-$ sign and spacelike for the $+$ sign. The $-$ sign in Eqs. (\ref{trans1}) and (\ref{trans2}) is the original sign of standard mimetic gravity \cite{Chamseddine:2013kea} however the $+$ sign is a generalization of the mimetic gravity.
 An amended type of mimetic gravity can process
the cosmological singularities \cite{Chamseddine:2016uef} and the singularity in the core of a black hole \cite{Chamseddine:2016ktu}. Furthermore, the initial attempt of the mimetic theory provides a guarantee that gravitational waves (GW) can travel at the speed of light, thereby supporting the consistency observed in recent findings such as the event GW170817 and its corresponding optical counterpart  \cite{Casalino:2018tcd,Casalino:2018wnc,Sherafati:2021iir}. Moreover, mimetic theory can investigate
the flat rotation curves of spiral galaxies without the need of dark matter \cite{Vagnozzi:2017ilo,Sheykhi:2019gvk}. From a cosmological point of view, the theory of mimetic has discussed a lot of interesting research papers in the past few years \cite{Chamseddine:2014vna,Baffou:2017pao,Dutta:2017fjw,Sheykhi:2018ffj,Abbassi:2018ywq,Matsumoto:2016rsa,Sebastiani:2016ras,Sadeghnezhad:2017hmr,Gorji:2019ttx,Gorji:2018okn,Bouhmadi-Lopez:2017lbx,Gorji:2017cai,Chamseddine:2019bcn,
Russ:2021ede,deCesare:2020swb,Farsi:2022hsy,Cardenas:2020srs,HosseiniMansoori:2020mxj,Arroja:2017msd} and  black
holes physics
\cite{Gorji:2020ten,Myrzakulov:2015sea,Nashed:2016tbj,Nashed:2011fg,Myrzakulov:2015kda,Ganz:2018mqi,Chen:2017ify,Nashed:2018aai,BenAchour:2017ivq,Brahma:2018dwx,Zheng:2017qfs,Zheng:2017qfs,Nashed:2018qag,Chamseddine:2019pux,Gorji:2020ten,
Sheykhi:2020fqf,Sheykhi:2020dkm,Nashed:2021ctg,Chamseddine:2021xhw,Bakhtiarizadeh:2021pyo,Nashed:2021pkc}. { It has been shown that for a spherically symmetric spacetime the only solution is the Schwarzschild spacetime which means that the Birkhoff's theorem is hold.} Moreover, mimetic theory have  been extended  to $f(R)$ gravity \cite{Nojiri:2014zqa,Odintsov:2015wwp,Oikonomou:2016pkp,Oikonomou:2016fxb,Oikonomou:2015lgy,Myrzakulov:2016hrx,Odintsov:2015cwa,Odintsov:2016imq,Odintsov:2016oyz,
Nojiri:2017ygt,Odintsov:2018ggm,Bhattacharjee:2021kar,Kaczmarek:2021psy,Chen:2020zzs}  and to Gauss-Bonnet gravitational theory \cite{Astashenok:2015haa,Oikonomou:2015qha,Zhong:2016mfv,Zhong:2018tqn,Paul:2020bje}. More specifically, a unified scenario   of early inflation and late-time acceleration in the framework of mimetic $f(R)$ gravity
was formulated  in \cite{Nojiri:2016vhu}. Moreover, it was assured that in the
frame of mimetic $f(R)$  gravity, the inflationary epoch
can be discussed \cite{Nojiri:2016vhu}.  In the present  study we discuss the interior of spherically symmetric solution within mimetic gravitational theory\footnote{Here in this study we will take the sign of Eq. (\ref{trans2}) as the one raised in the original mimetic theory.}.  Because of the non-trivial contribution of the mimetic field $\zeta$ the Einstein field equation gives:
 \begin{equation}\label{2}
{G_{\mu\nu} +\frac{1}{2} g_{\mu\nu}(\partial^\mu\zeta \partial^\nu \zeta+1)=\kappa T_{\mu\nu}}\,,
\end{equation}
{ where $T_{\mu\nu}$ is the energy-momentum tensor and $\kappa=\frac{8\pi\,G}{c^4}$ is the gravitational constant and $G_{\mu \nu}$ is the Einstein tensor defined as:}
\begin{align}
{ G_{\mu \nu}=R_{\mu \nu}-\frac{1}{2}G_{\mu \nu} R}\,, \end{align}
{ where $R_{\mu \nu}$ is the Ricci tensor defined as:}\[{ R_{\mu \nu}={R^\alpha}_{\mu \alpha \nu}={\Gamma^\alpha}_{\mu \nu,\alpha}-{\Gamma^\alpha}_{\mu \alpha,\nu}+{\Gamma^\alpha}_{\nu \mu}{\Gamma^\beta}_{\alpha \beta}-{\Gamma^\alpha}_{\nu \beta}{\Gamma^\beta}_{\alpha \mu}}\,,\]
{ with ${\Gamma^\alpha}_{\nu \mu}$ being the Christoffel symbols second kind and ${R^\alpha}_{\mu \alpha \nu}$ is the Riemann tensor fourth order and $R$ is the Ricci scalar defined as $R=g^{\mu \nu}R_{\mu \nu}$.  Equations (\ref{2}) coincides with Einstein GR when the scalar field has a constant value, i.e., $\partial^\mu\zeta \partial^\nu \zeta+1=0$. There are many applications in the framework of mimetic in cosmology as well as in solar system \citep[see for example][]{Saadi:2014jfa,Haghani:2015iva,Addazi:2017cim,Astashenok:2015qzw,Babichev:2016jzg}}.

The current study is structured as follows: In Section \ref{S3}, we utilize mimetic field equations, specifically equation (\ref{2}), to analyze a spherically symmetric object with an anisotropic matter source. This results in a system of three nonlinear differential equations with five unknown functions, including two metric potentials, energy density, radial pressure, and tangential pressure. To close the system, we impose two additional constraints: we assume a specific form for one of the metric potentials, $g_{tt}$, which is commonly done in interior solutions, and we assume the vanishing of anisotropy and derive the form of the spatial component of the metric potential, $g_{rr}$. Collecting this information, we obtain the analytic expressions for the energy density and pressure that satisfy the mimetic equation of motion. In subsection \ref{S4}, we delineate the physical requirements that any isotropic stellar model must meet to be in agreement with a genuine compact star. In Section \ref{S5}, we discuss the applicability of the derived solution under the conditions presented in Section \ref{S4}. In Section \ref{S6}, We integrate our model using the Schwarzschild solution, an external vacuum solution, and make adjustments the model parameters based on the properties of the pulsar $\textit{Cen X-3}$, which has a mass estimate of $M= 1.49 \pm 0.49, M_\odot$ and a radius of $R= 9.178 \pm 0.13$ km. In Section \ref{S7}, we investigate the stability of the model using the TOV equation of hydrostatic equilibrium and the adiabatic index. Finally, we summarize our findings in Section \ref{S8}.

\section{spherically symmetric interior solution  }\label{S3}
{ To be able to derive an interior solution we will use a spherically symmetric spacetime to make the calculations and discussion more easy. For this aim, we assume  the spacetime of spherical symmetric to  have the form:
 \begin{eqnarray} \label{met12}
& &  ds^2=E^2(r)dt^2-\frac{1}{E_1(r)}\,dr^2-r^2(d\theta^2-\sin^2\theta d\phi^2)\,,  \end{eqnarray}
with $E(r)$ and $E_1(r)$  are  unknown  functions. When $E=E_1$ one can recover Schwarzschild solution for exterior Einstein GR.} Using Eq. (\ref{met12}), we get the Ricci tensor and Ricci scalar in the form:
  \begin{eqnarray} \label{Ricci}
  &&{ {\cal R}_{tt}(r)=\frac{E(2rE_1E''+E'[rE'_1+4E_1])}{2r}}\,,\nonumber\\
  &&{ {\cal R}_{rr}(r)=-\frac{2rE_1E''+E'_1[rE'+2E])}{2EE_1r}}\,,\nonumber\\
  &&{  {\cal R}_{\theta \theta}(r)=-\frac{2rE_1E'+rEE'_1-2E(1-E_1)}{2E}\,,\qquad {\cal R}_{\phi \phi}(r)={\cal R}_{\theta \theta}(r)\sin^2\theta}\,,\nonumber\\
&&{ {\cal R}(r)=-\frac {E' E'_1 {r}^{2}+2\,E_1 E''
{r}^{2}+4\,E_1 E' r+2\,E'_1 E r-2\,E +2\,EE_1 }{E {r}^{2}}}\,,
  \end{eqnarray}
  where $E\equiv E(r)$,  $E_1\equiv E_1(r)$,  $E'=\frac{dE}{dr}$, $E''=\frac{d^2E}{dr^2}$ and $E_1'=\frac{dE_1}{dr}$.
  Plugging Eq.
 (\ref{2}) with Eq. (\ref{met12}) and by  using Eq. (\ref{Ricci}) we get:

 \begin{eqnarray}
&& \textrm{ The t\,t component of mimetic field equation is:}\nonumber\\
&&\rho=\frac {1-rE'_1-E_1}{r^2}\,,\nonumber\\
&& \textrm{ The r\,r component of mimetic field equation is:}\nonumber\\
&&p={\frac {2\,\zeta'^{2} E_1{}^{2}{r}^{2}E'' +\left[r \left( 4\,E_1  + E'_1 r \right) E'  +E   \left(2E_1 -2+{r }^{2}\rho  -3{r}^{2}p +2 E'_1r \right)  \right] E_1  \zeta'^{2}+2E' rE_1  +E   \left(E_1-1   \right) }{E {r}^{2}}}\,,\nonumber\\
&& \textrm{ The}\, \mathrm{\theta\, \theta = \phi \, \phi}\textrm{ component of Mimetic field equation is:}\nonumber\\
&&p_1=\frac{2rE_1E''+2E_1E'+E'_1(E+rE')}{2E {r}}\,,\nonumber\\
\label{feq}
\end{eqnarray}
{ where we have  set the Einstein gravitational constant, i.e., $\kappa$, to unity.
For an anisotropic fluid with spherical symmetry, we assume the energy-momentum tensor., i.e.}
\begin{equation}\label{Tmn-anisotropy}
{    {T}{^\alpha}{_\beta}=(p_{1}+\rho)u{^\alpha}u{_\beta}+p_{1} \delta ^\alpha _\beta + (p-p_{1})\chi{^\alpha}\chi{_\beta}\,.}
\end{equation}
{ Here, $\rho=\rho(r)$ represents the energy density of the fluid, $p=p(r)$ denotes its radial pressure  and $p_{1}=p_{1}(r)$ represents the tangential pressure. As a result, the energy-momentum tensor takes the form ${T}{^\alpha}{\beta}=diag(-\rho,\,p,\,p{1},\,p_{1})$. If the mimetic scalar field has a constant value, or $\zeta=C$, then equations (\ref{feq}) will be equivalent to the interior differential equations of Einstein's general relativity \cite{Nashed:2020buf,Nashed:2020kjh}..} The differential equations (\ref{feq}) are three non-linear  in six unknowns $E$, $E_1$, $\rho$, $p$, $p_1$ and the mimetic field $\zeta$ which we can fix it form the use of Eq. (\ref{trans2}), i.e.,\[\zeta=\frac{1}{\sqrt{-E_1}}.\] Therefore, to put the above system in a solvable form we  need two extra conditions. The first one is to suppose the temporal component of the metric potential $E$ in the form \cite{Torres-Sanchez:2019wjv,Estevez-Delgado:2018bxa}:
\begin{equation}\label{metg}
E(r)={\frac {a_0 \left( 5+4\,a_1{r}^{2} \right) }{\sqrt {1+a_1{r}^{2}}}}\,,
\end{equation}
where $a_0$ is a constant that has no dimension and $a_1$ is another constant that has  dimension of inverse length square, i.e., $L^{-2}$.
The second condition is the use of r\,r and  $\mathrm{\theta\, \theta}$ components of Eq. (\ref{feq}), i.e., the anisotropy equation,   and  imposing  of Eq. (\ref{metg}) yields:

\begin{align}\label{metf}
&E_1(r)=  \frac{\left( 1+2\,a_1{r}^{2}+{a_1}^{2}{r}^{4} \right)}{\varepsilon^{3}}\biggl\{[5+6\,a_1 r^2]\varepsilon- 4\,a_1[1+a_1{r}^{2}] {r}^{2}\varepsilon_1+a_2{r}^{2}+a_2\,a_1{r}^{4} \biggr\} \,.
\end{align}
{ Here $a_2$ is a constant of integration  with inverse length square dimension, i.e., $L^{-2}$, $\varepsilon=\sqrt {5+12\,a_1{r}^{2}+8\,{a_1}^{2}{r}^{4}}$ and $\varepsilon_1={\arctanh} \left( {\frac {1+2\,a_1{r}^{
2}}{\varepsilon}} \right)$.}

Using Eqs. (\ref{metg}) and (\ref{metf}) in the system of differential Eqs. (\ref{feq}),  we obtain the components of the energy-momentum  in the form:
\begin{eqnarray}\label{sol}
&&\rho={\frac {1}{\varepsilon^6}}\, \biggl\{ 12\, \left( 1+a{r}^{2} \right) ^{2} \left(  \left( 3a_1{r}^{2}+1 \right) \varepsilon^{3}-4 \left( 3+4a_1{r}^{2} \right)  \left( 1+ a_1{r}^{2} \right) \varepsilon a_1{r}^{2} \right) a_1\varepsilon_1-3\, \left(1+3a_1{r}^{2} \right) \left( 1+a_1{r}^{2} \right) ^{2}a_2\, \varepsilon^{3}\nonumber\\
&& + \biggl(12  \left( 3+4a_1{r}^{2 } \right)  \left( 1+a_1{r}^{2} \right) ^{3}{r}^{2}a_2\,\varepsilon+ \varepsilon  \left( 144{a_1}^{4}{r}^{8}+424\,{a_1}^{3}{r}^{6}+486\,{a_1}^{2}{r}^{4}+265\,a_1{r}^{2}+60 \right)  \biggr) a_1 \biggr\}  \,,\nonumber\\
&&p={\frac {1}{\varepsilon  ^{6} \left( 5+4a_1{r}^{2} \right)}}\, \biggl\{  \left(36 {a_1}^{2}{r}^{4}+33\,a_1{r}^{2}+5 \right)  \left( 1+a_1{r}^{2} \right) ^{2}a_2\, \varepsilon^{3}-4 \left( 1+a{r}^{2} \right) ^{2}a_1\left( 12{a_1}^{2}{r}^{4}+15\,a_1{r}^{2}+5 \right) \varepsilon^{3}\varepsilon_1\nonumber\\
&&\nonumber\\
&&+\varepsilon^{3} \biggl( \varepsilon^{1/2} \left( 72{a_1}^{3}{r }^{6}+190\,{a_1}^{2}{r}^{4}+167\,a_1{r}^{2}+50 \right) - 6\left( 3+4a_1{r}^{2} \right)  \left( 1+a_1{r}^{2 } \right) ^{2}{r}^{2}a_2\, \biggr) a_1 \biggr\} \,.\nonumber\\
&&
\end{eqnarray}
{ The energy density of Eq. (\ref{sol}) is the same as of GR for isotropic solution \cite{Torres-Sanchez:2019wjv} however, the pressure is different. This difference is due to the contribution of the mimetic scalar field. It should be noted that if the mimetic scalar field is set equal zero in Eq. (\ref{feq}) and solving the system using ansatz (\ref{metg}) we get the form of density and pressure presented in \cite{Torres-Sanchez:2019wjv}. Moreover, it is important to stress that the use of metric potentials (\ref{metg}) and (\ref{metf}) in the system (\ref{feq}) gives $p=p_1$ which insure the isotropy of our model.}

The mass contained in a sphere that has radius $r$  is given by:
\begin{align}\label{mas}
M(r)=4\pi {\int_0}^r \rho(\eta) \eta^2 d\eta\,.\end{align}
By employing the expression for energy density provided in Equation (\ref{sol}) and substituting it into Equation (\ref{mas}), we obtain the asymptotic representation of the mass as:
\begin{eqnarray}\label{mas1}
&&M(r)\approx
 (-0.3139182118a_1-0.04472135955a_2)r^3+a_1(0.08835092710a_1{}^2+0.02683281574a_1a_2)r^5 \nonumber\\
&&\nonumber\\
&&-a_1{}^2(0.0883509271a_1+0.02683281574a_2)r^7+a_1{}^3(0.03114573467a_1+0.0196773982a_2)r^9
\,.\end{eqnarray}
The compactness parameter  with radius $r$  of a spherically symmetric
source is defined as \cite{Singh:2019ykp,Roupas:2020mvs}:
\begin{eqnarray}\label{gm1}
&&C(r)=\frac{2M(r)}{r}.
\end{eqnarray}

In the next  subsection, we present the physical conditions  that are viable for an isotropic stellar structure and examine if model  (\ref{sol}) satisfy  them or not.
{ \subsection{Necessary  criteria  for a physically viable  stellar isotropic model}}\label{S4}
Before we proceed we are going to use the following dimensionless substitution: \[r=xR\,,\] where $R$  is the radius of the star and $x$ is a dimensionless constant that equal to one when $r=R$ and equal zero at the center of the star. Also we assume the dimensional constants $a_1$ and $a_2$ to take the form: \begin{align}\label{con11} a_1=\frac{u}{R^2}\,, \qquad a_2=\frac{w}{R^2}.\end{align} where $u$ and $w$ are  dimensional quantities.   By using the substitution of $a_1$, $a_2$  and $r$ into the physical components of model, Eqs. (\ref{metg}), (\ref{metf}) and (\ref{sol}) we will get a dimensionless physical quantities. Now we are ready to discuss the necessary criteria that we apply in the isotropic model:

A physical isotropic model must verify:\vspace{0.1cm}\\
$\bullet$ The  metric potentials $E(x)$ and $E_1(x)$, and  $\rho$ and $p$ must  have good behavior  at the
core of the stellar object and have  regular behavior  through the structure of the star without singularity.\vspace{0.1cm}\\
$\bullet$  The energy density component, denoted as  $\rho$, is required to be positive within the internal structure of the star. Additionally, it should possess a finite positive value and exhibit a monotonically decreasing trend towards the surface of the stellar interior, i.e.,  $\frac{d\rho}{dx}\leq 0$.\vspace{0.1cm}\\
$\bullet$  The pressure, denoted as $p$, must maintain a positive value throughout the fluid structure, meaning $p\geq0$. Furthermore, the derivative of pressure with respect to the spatial variable,  i.e., $\frac{dp}{dx}< 0$, must be, indicating a decreasing pressure gradient. Additionally, at the surface, $x = 1$, (corresponding to $r = R$), the pressure $p$ should be zero.\vspace{0.1cm}\\
$\bullet$   The  energy conditions of  isotropic star requires the following
inequalities:\vspace{0.1cm}\\
(i)The condition of weak energy (WEC):  $p+\rho > 0$, $\rho> 0$. \hspace{0.1cm}
(ii) The conditions of dominant energy (DEC): $\rho\geq \lvert p\lvert$. \hspace{0.1cm}\\
(iii) The condition of strong energy  (SEC): $p+\rho > 0$, $\rho+3p > 0$.  \vspace{0.1cm}\\
$\bullet$ The causality condition must be verified, to  have a viable  true
model, i.e., $v<1$ where $v$ is the speed of sound.\vspace{0.1cm}\\
$\bullet$  The interior metric potentials, $E$ and $E_1$,   must be   joined smoothly to
the exterior metric potentials (Schwarzschild metric) at the
surface of the stellar, i.e., $x=1$.\vspace{0.1cm}\\
$\bullet$ For a  true  star the adiabatic index is
greater than $\frac{4}{3}$.\vspace{0.1cm}\\

Now, we are ready to examine the above-listed physical criteria on our model to see if it satisfies all of them or not.

\section{The Physical behaviors  of model (\ref{sol})}\label{S5}
\subsection{The free singularity of the model}
a- The metric potentials given by Eqs (\ref{sol}) and (\ref{metf}) fulfill:
\begin{align}\label{sing}
E_{x\rightarrow 0}=25a_0{}^2\,\qquad  \qquad \textrm{and} \qquad \qquad {E_1}_{x\rightarrow 0}=1\,.
\end{align}
Equation (\ref{sing}) guarantees that the lapse functions possess finite values at the core of the stellar configuration. Additionally, the derivatives of the metric potentials with respect to x must also have finite values at the core, i.e., $f'(x=0)=f'_1(x=0)=0$. Equations (\ref{sing}) ensures that the laps functions  are regular at the core and have good behavior throughout the center of the star.\vspace{0.1cm}\\
{  ii-The density and  pressure of Eq. (\ref{sol}) take the following form at  core:}
\begin{align}\label{reg}
&{ \rho_{_{_{x\rightarrow 0}}}=\frac {12\,u\sqrt {5}{\arctanh} \left( 1/\sqrt{5} \right)-3\,w\sqrt {5}-
60\,u}{}{25R}^{2}}\,, \nonumber\\
&{  {p}_{_{_{x\rightarrow 0}}}=\frac {50\,u-4\,u\sqrt {5}{\arctanh} \left( 1/\sqrt{5} \right)+w\sqrt {5}
}{{25R}^{2}}}\,.
\end{align}
{  Equation (\ref{reg})  ensures  the positivity of density and pressure  assuming}  \[{ 12\,u\sqrt {5}{\arctanh} \left( 1/\sqrt{5} \right)-3\,w\sqrt {5}-
60\,u>0\,,\qquad   {\textrm and} \qquad 50\,u-4\,u\sqrt {5}{\arctanh} \left( 1/\sqrt{5} \right)+w\sqrt {5}>0}\,.\]
{  Moreover, the Zeldovich condition \cite{1971reas.book.....Z} that connects the  density and pressure at the center of the star through the inequality, i.e.,  $\frac{p(0)}{\rho(0)}\leq 1$. Applying  Zeldovich condition in Eq. (\ref{reg}), we get:}
\begin{align}\label{reg1}
{  \frac {50\,u-4\,u\sqrt {5}{\arctanh} \left( 1/\sqrt{5} \right)+w\sqrt {5} }{12\,u\sqrt {5}{\arctanh} \left( 1/\sqrt{5} \right)-3\,w\sqrt {5}-
60\,u}\leq 1}\,,\end{align}
{  which yields:
$$\Rightarrow w\leq\frac{\left[8\sqrt {5}{\arctanh}\left( 1/\sqrt{5} \right)- 55\right]u}{2\sqrt{5}}\,.$$}
iii-The derivatives  of density, $\rho$,  and
pressure, $p$, of Eq.  (\ref{sol}) are respectively:
\begin{eqnarray}\label{dsol}
&&\rho'=-\frac{2\,x{u}^{2}}{ {R}^{2}\varepsilon^{7/2}} \biggl\{ 192\,w\,{u}^{5}{x}^{10}+192\,\varepsilon_1 {u}^{5}{x}^{10}+1248\,\varepsilon_1 {u}^{4}{x}^{8}-32\,\varepsilon {x}^{4}{u}^{4}{x}^{8}+1248\,w\,{u}^{4}{x}^{8}-96\,\varepsilon{u }^{3}{x}^{6}+3048\,{w}\,{u}^{3}{x}^{6}\nonumber\\
 &&+3048\,\varepsilon_1 {u}^{3}{x}^{6}-372\,\varepsilon{u}^{2}{x}^{4}+3372\,\varepsilon_1 {u}^{2}{x}^{4}+3372 \,w\,{u}^{2}{x}^{4}-480\,\varepsilon\,u{x}^{2}+1680\,\varepsilon_1 u{x}^{2}+1680\,w\,u{x }^{2}+300\,w+300\,\varepsilon_1 -175\,\varepsilon \biggr\}
\,,
  \nonumber\\
 && p'=\frac{2\,xw}{\left( 5+4\,u{x}^{2} \right) ^{2 }{R}^{2} \varepsilon^{5/2} } \biggl\{ 96\,w\,{u}^{5}{x}^{10}-384\,{w}^{6}\varepsilon_1 {x}^{10}+64\,{u}^{5}\zeta{x}^{8}-1792\,{u}^{5}\varepsilon_1{x}^{8}+448\,w\,{u}^{4}{x}^{8}+160\,{u}^{4}\varepsilon{x}^{6}-3448\,{u}^{4}\varepsilon_1 {x}^{6}\nonumber\\
 &&+862\,w\,{u}^{3}{x}^{6}-3340\,{u}^{3}\varepsilon_1 {x}^{4}+244 \,{u}^{3}\varepsilon{x}^{4}+835\,w \,{u}^{2}{x}^{4}-1600\,{u}^{2}\varepsilon_1 {x}^{2}+220\,{u }^{2}\varepsilon{x}^{2}+400\,w\,u{ x}^{2}-300\,u\varepsilon_1+75\,u\varepsilon+75\,w \biggr\}  \,,\nonumber\\
 &&
\end{eqnarray}
where $\rho'=\frac{d\rho}{dx}$ and  $p'=\frac{dp_r}{dx}$.  Fig. \ref{Fig:2}  \subref{fig:grad} shows that the gradients  of the components of energy-momentum tensor behave in negative way.\vspace{0.1cm}\\
iv-The speed  of sound (when c = 1) yields:
\begin{eqnarray}\label{dso2}
&&v{}^2=\frac{dp}{d\rho}=\frac{\varepsilon^2}{\left( 5+4\,u{x}^{2} \right) ^{2}} \bigg\{96\,w\,{u}^{5}{x}^{10}-384\,{u}^{6}\varepsilon_1{x}^{10}+64\,{u}^{5}\varepsilon{x}^{8}-1792\,{u}^{5}\varepsilon_1 {x}^{8}+448\,w\,{u}^{4}{x}^{8}+160\,{u}^{4}\varepsilon {x}^{6}-3448\,{u}^{4}\varepsilon_1 {x}^{6}\nonumber\\
 &&+862\,w\,{u}^{3}{x}^{6}-3340\,{u}^{3}\varepsilon_1{x}^{4}+244 \,{u}^{3}\varepsilon{x}^{4}+835\,w \,{u}^{2}{x}^{4}-1600\,{u}^{2}\varepsilon_1{x}^{2}+220\,{u }^{2}\varepsilon{x}^{2}+400\,w\,u{ x}^{2}-300\,u\varepsilon_1 +75\,u\varepsilon+75\,w \bigg\}  \nonumber\\
 &&\bigg\{48\,w\,{u}^{5}{x}^{10} -192\,{u}^{6 }\varepsilon_1{x}^{10}+32\,{ u}^{5}\varepsilon {x}^{8}-1248\,{u}^{5}\varepsilon_1 {x}^{8}+312\,w\,{u}^{4}{x}^{8}+96\,{u}^ {4}\varepsilon{x}^{6}-3048\,{u}^{4}\varepsilon_1{x}^{6}+75\,w\nonumber\\
 &&+762\,w\,{u}^{3}{x}^{6}-3372\,{u}^{3 }\varepsilon_1 {x}^{4}+372\,{u}^{3}\varepsilon{x}^{4}+843\,w\,{u}^{2}{x}^{4}+480\,{u}^{2} \varepsilon{x}^{2}-1680\,{u}^{2}\varepsilon_1 {x}^{2}+420\,w\,u{x}^{2}+175\,u\varepsilon-300\,u\varepsilon_1 \biggr\}^{-1}
\, .\nonumber\\
 &&
\end{eqnarray}
which is less than unity as Fig. \ref{Fig:2}  \subref{fig:sped} shows.
\subsection{Junction conditions}
We make the assumption that the exterior solution of the star is a vacuum, described by the Schwarzschild solution. This is because, in the mimetic theory, the Schwarzschild solution is the only exterior spherically symmetric solution \cite{Nashed:2020buf,Nashed:2020kjh}. The form of the Schwarzschild solution is given by\footnote{{The isotropic Schwarzschild solution is given by \cite{Shirafuji:1995xc} \begin{align}\label{met11}
ds^2=\frac{(1-M/2r)^2}{(1+M/2r)^2}dt^2-\frac{1}{(1+M/2r)^4}[dr^2+r^2(d\theta^2+\sin^2\theta d\phi^2)]\,.\end{align}
The above metric is the one that we use in the junction conditions but due to the nature of the line-element (\ref{met12}), so it is logic to match it with the asymptotic form of line-element (\ref{met11}) which is give by:
\begin{align}\label{met111}
ds^2\approx \Big(1-\frac{2M}{r}\Big)dt^2-\Big(1+\frac{2M}{r}\Big)dr^2-r^2(d\theta^2+\sin^2\theta d\phi^2)\,.\end{align}}}:
\begin{eqnarray}\label{Eq1}  ds^2= -\Big(1-\frac{2M}{xR}\Big)dt^2+\Big(1-\frac{2M}{xR}\Big)^{-1}dr^2+x^2R^2d\Omega^2,
 \end{eqnarray}
where $M$ is the mass of the star.   The junction of the laps
functions at   $x=1$  gives:
\begin{eqnarray}\label{Eq2}  E(x\rightarrow1)=\left(1-\frac{2M}{R}\right), \qquad \qquad E_1(x\rightarrow1)=\left(1-\frac{2M}{R}\right)^{-1}\,,
 \end{eqnarray}
in addition to the constrain of the vanishing of pressure at the surface we fix  the dimensionless constants of solution  (\ref{metg}) and (\ref{metf}) as:
\begin{eqnarray}\label{Eq3} &&a_0=\pm{\frac {\sqrt {R \left( R+R{u}^{2}-2\,M-2\,M{u}^{2} \right) }}{R \left( 5+4\,{u}^{2} \right) }}
\,,\nonumber\\
&&u=-\frac {48\,\sqrt {2}M+R\left[19\sqrt {2}-9\,R\,\varrho\right]-\sqrt{ M\left[4608\,{M}-864\,\sqrt {2}R +3648R\right]+[351\,\varrho^{2} -1482\,
\varrho \sqrt {2}+2702]\,{R}^{2}}}{
24R \left(3\,\sqrt {2} -\varrho  \right) }\,,\nonumber\\
&&w={\frac {u \left[4\left( u+1 \right) ^{ 2} \left(12 {u}^{2}+15u+5 \right)\varepsilon_1- \left(72 {u}^{3}+ 190\,{u}^{2}+167\,u+50 \right) \varepsilon  \right]}{\left( 47\,{u}^{2}+39\,{u}^{3}+12\,{u}^{4}+25\, u+5 \right) }}  \,,\nonumber\\
 \end{eqnarray}
 where $\varrho= \arctanh \left(\frac{ 1}{\sqrt {2}}  \right)$.
 \section{Examination of the  model (\ref{sol}) with true compact stars}\label{S6}
Now, we are ready to use the previously listed  conditions in Eq. (\ref{sol}) to examine the physical masses and radii of the  stars. To extract more information of  the model (\ref{sol}), we  use
the pulsar \textrm{Cen X-3} which has  mass   $M = 1.49\pm 0.49 M_\circledcirc$ and  radius $R = 9.178\pm0.13$ km, respectively \cite{Gangopadhyay:2013gha}.
In this study the  value of mass is  $M=1.98 M_\circledcirc$ and  the radius is $R=9.308km$. These conditions  fix the dimensionless constants $a_0$, $u$ and $w$ as\footnote{ {When the constants $a_0$, $a_1$ and $a_2$ equal zero then the density and pressure vanish and in that case we get a vacuum  solution which is the Schwarzschild solution.}}:
\begin{eqnarray}&&a_0=-0.1509026812\,,\qquad \qquad u=0.1241376545\,, \qquad w=-2.382227207\,.\end{eqnarray}

 Using the above values of constants we  plot the  physical quantities of the model (\ref{sol}).
\begin{figure}
\centering
\subfigure[~Energy-density given by Eq. (\ref{sol}) ]{\label{fig:dens}\includegraphics[scale=.3]{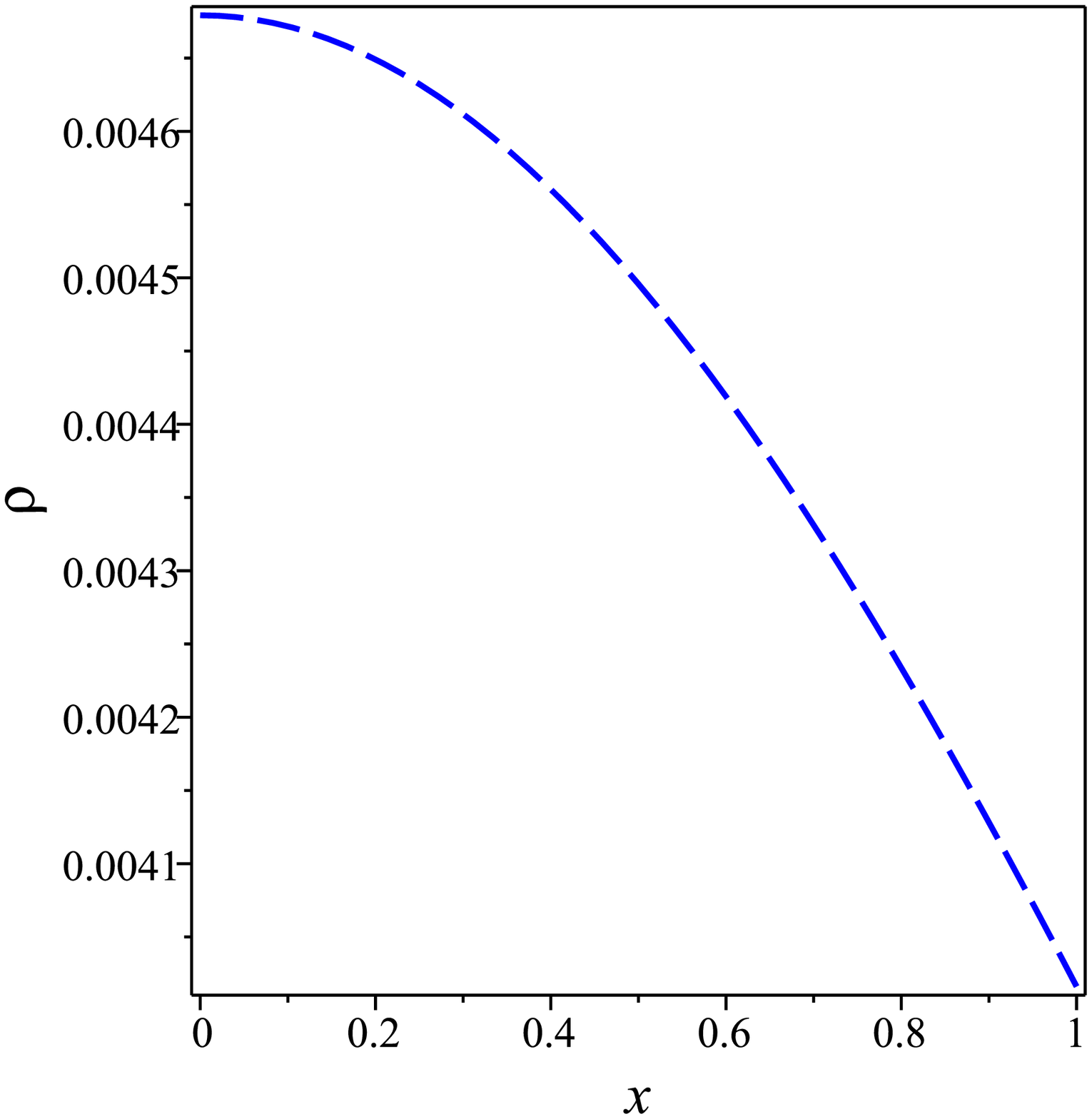}}
\subfigure[~ Pressure given by Eq. (\ref{sol})]{\label{fig:pressure}\includegraphics[scale=.3]{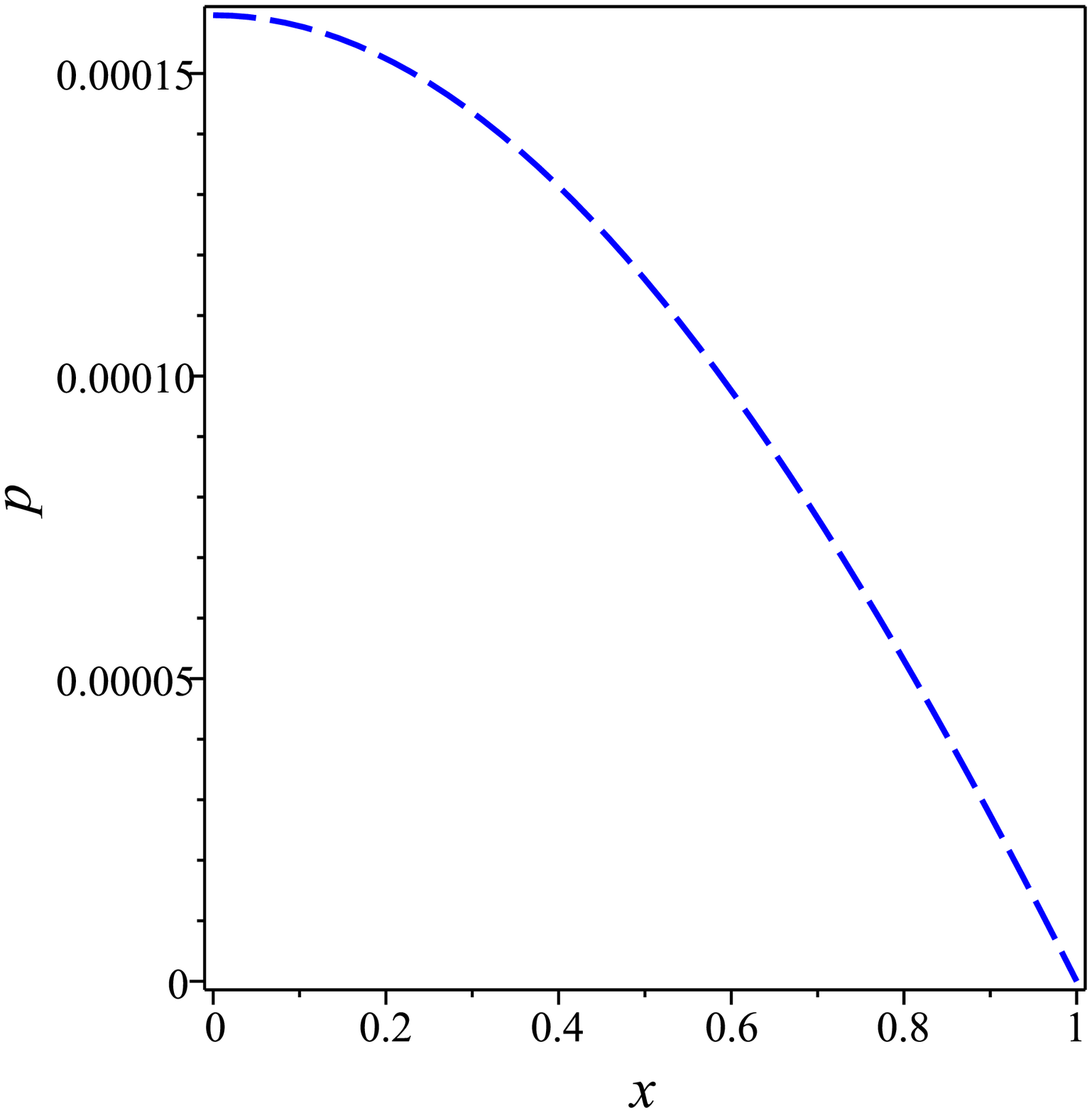}}
\caption[figtopcap]{\small{Plots of Fig. \subref{fig:dens} the density and Fig. \subref{fig:pressure} pressure of (\ref{sol})  versus the dimensionless   $x$   using the constants fixed from  \textrm{Cen X-3} \cite{Naik:2011qc}.}}
\label{Fig:1}
\end{figure}
In  Figs. \ref{Fig:1} \subref{fig:dens}, and  \subref{fig:pressure}  we depict  energy-density
 and   pressure  of
 the star \textrm{Cen X-3} which shows that density and pressure possess positive values as necessary for true stellar configuration moreover,  the density is high at the center and decreases toward the surface of the star. additionally, Fig.  \ref{Fig:1}  \subref{fig:pressure} shows that the value of the pressure is zero at the surface of the stellar. The behaviors of density and pressure presented in Figs. \ref{Fig:1} \subref{fig:dens}, and  \subref{fig:pressure} are appropriate  to a true model.
\begin{figure}
\centering
\subfigure[~Speed of sound]{\label{fig:grad}\includegraphics[scale=0.3]{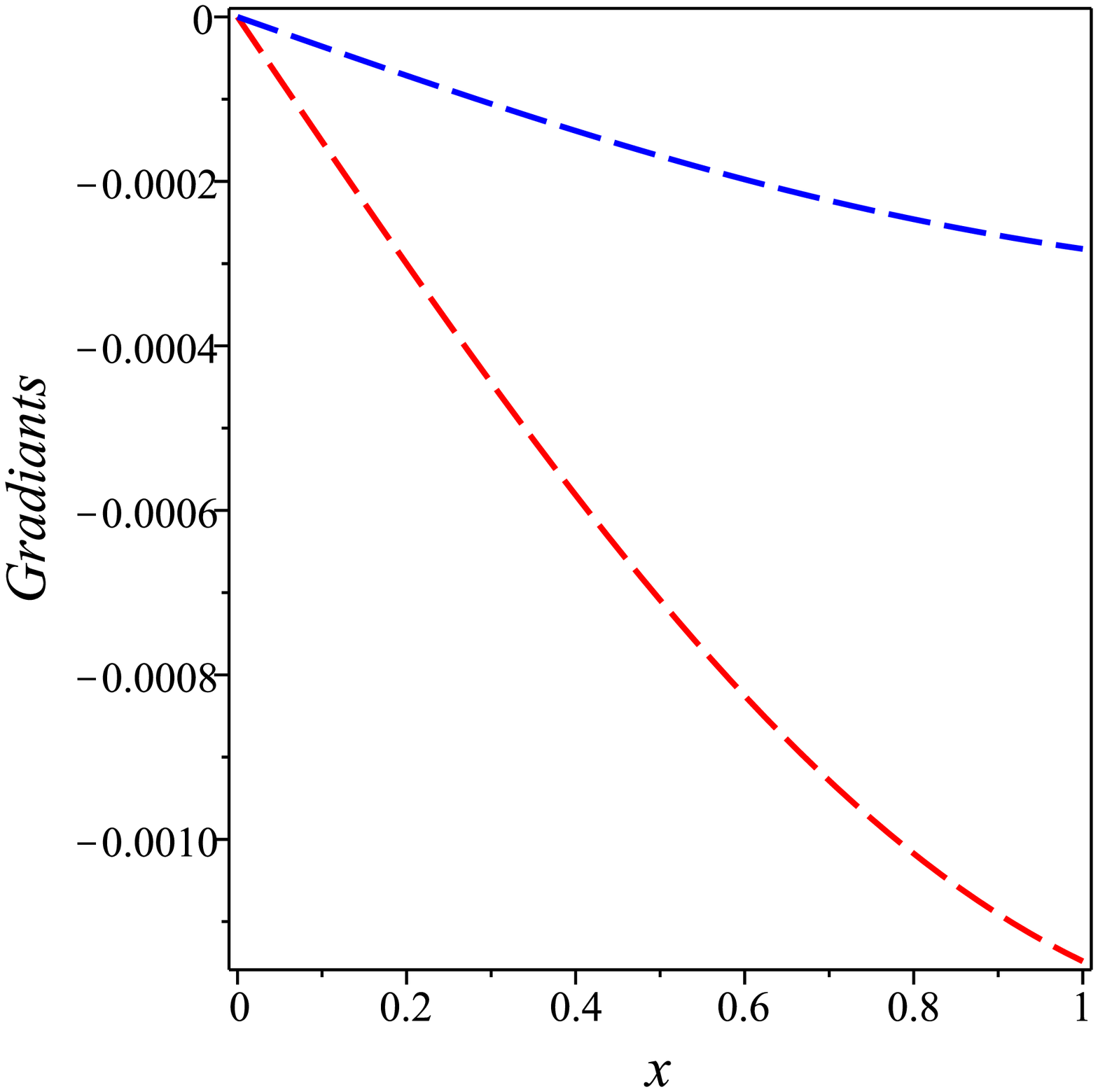}}
\subfigure[~Weak  energy conditions]{\label{fig:sped}\includegraphics[scale=0.3]{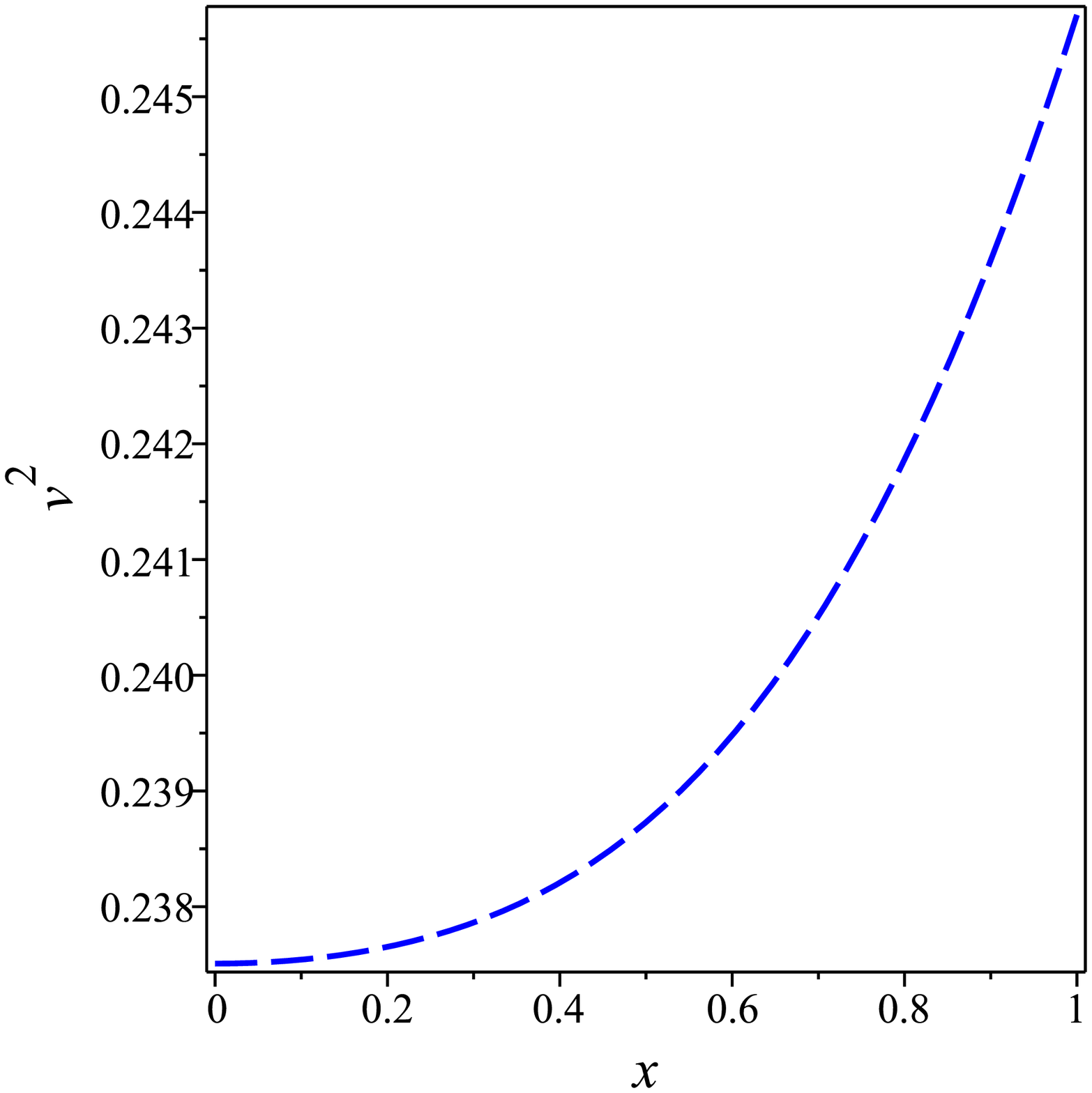}}
\subfigure[~Weak  energy conditions]{\label{fig:WEC}\includegraphics[scale=0.3]{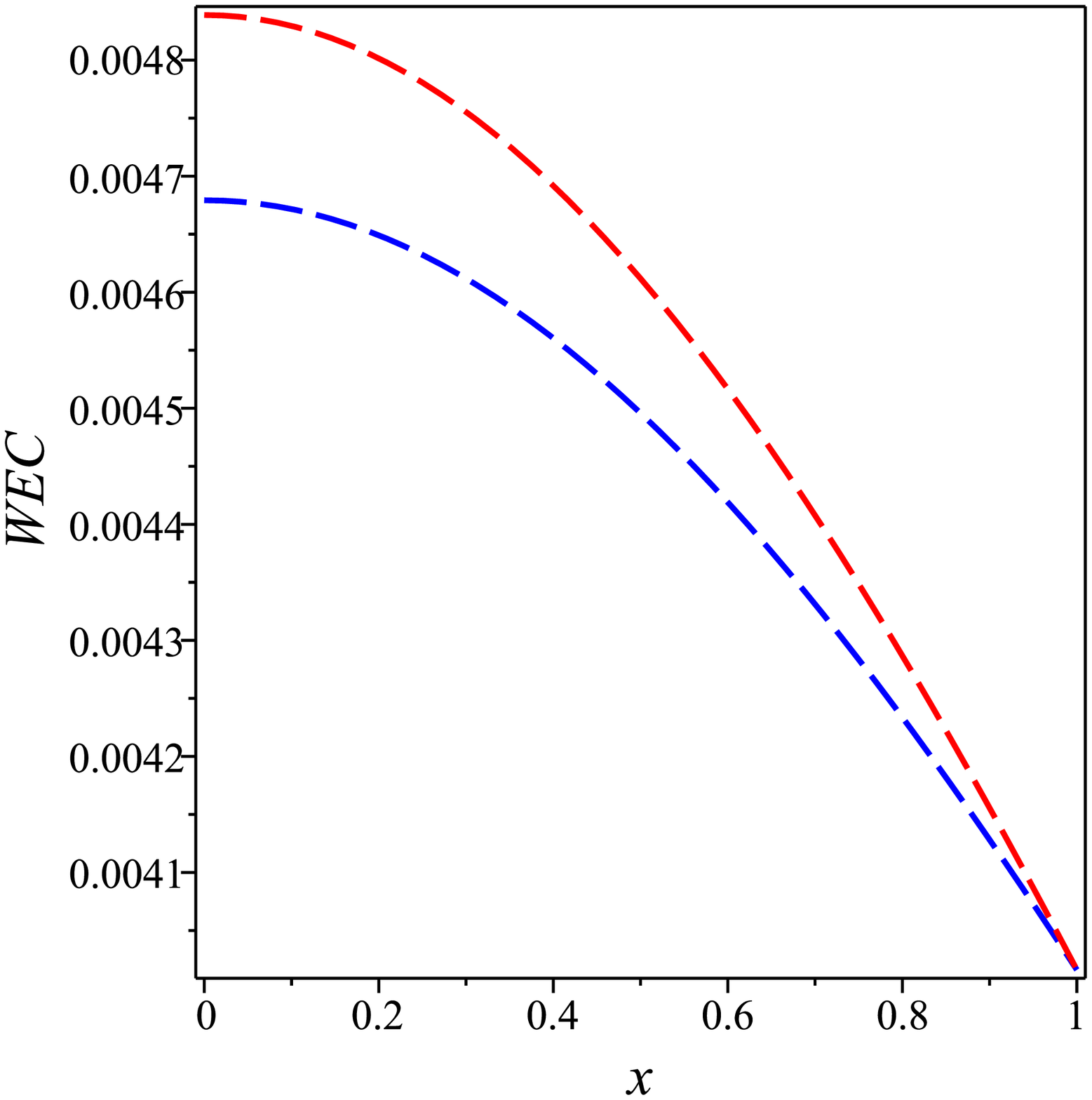}}
\subfigure[~Strong energy condition]{\label{fig:SEC0}\includegraphics[scale=.3]{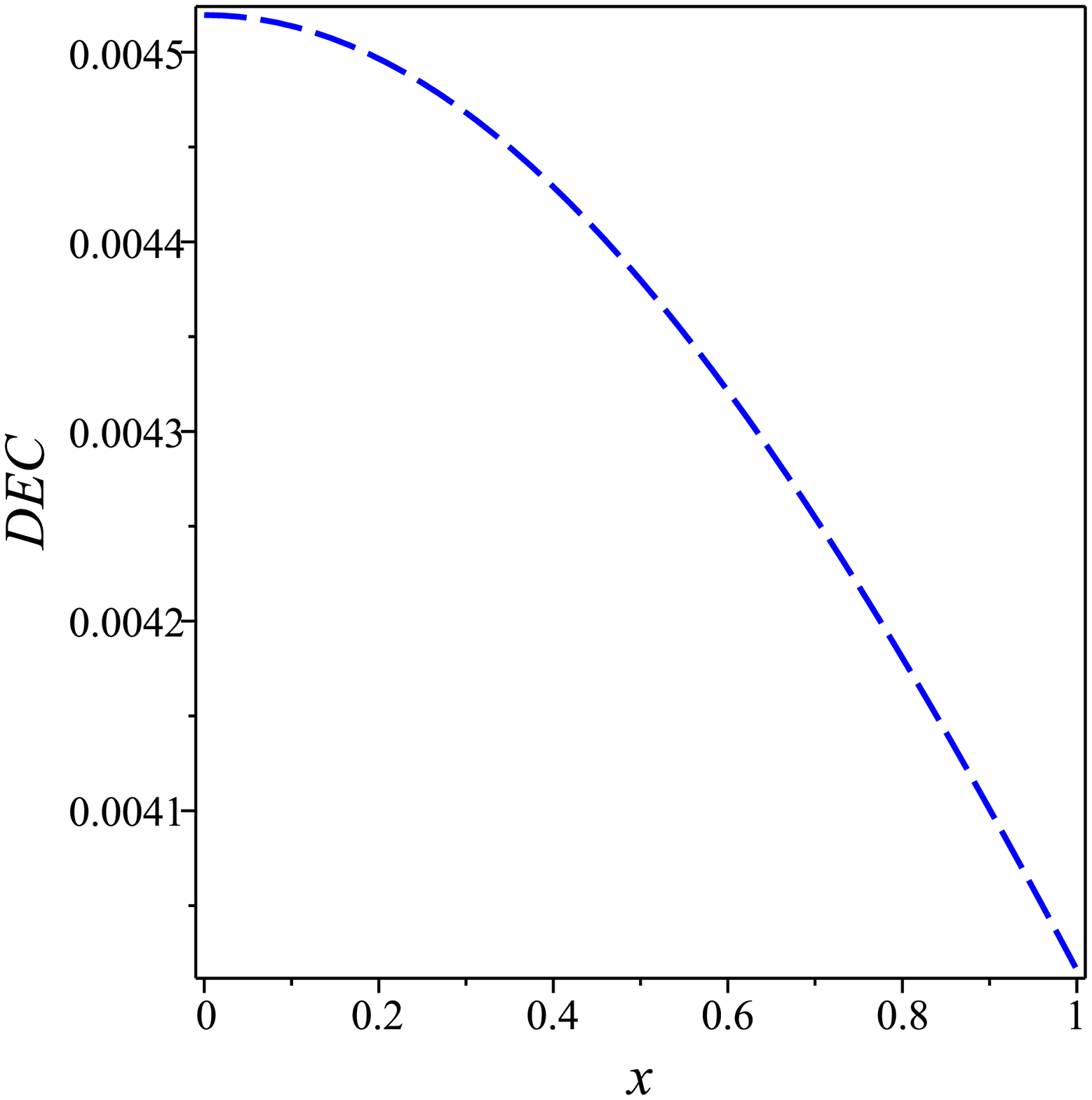}}
\subfigure[~Strong energy condition]{\label{fig:SEC}\includegraphics[scale=.3]{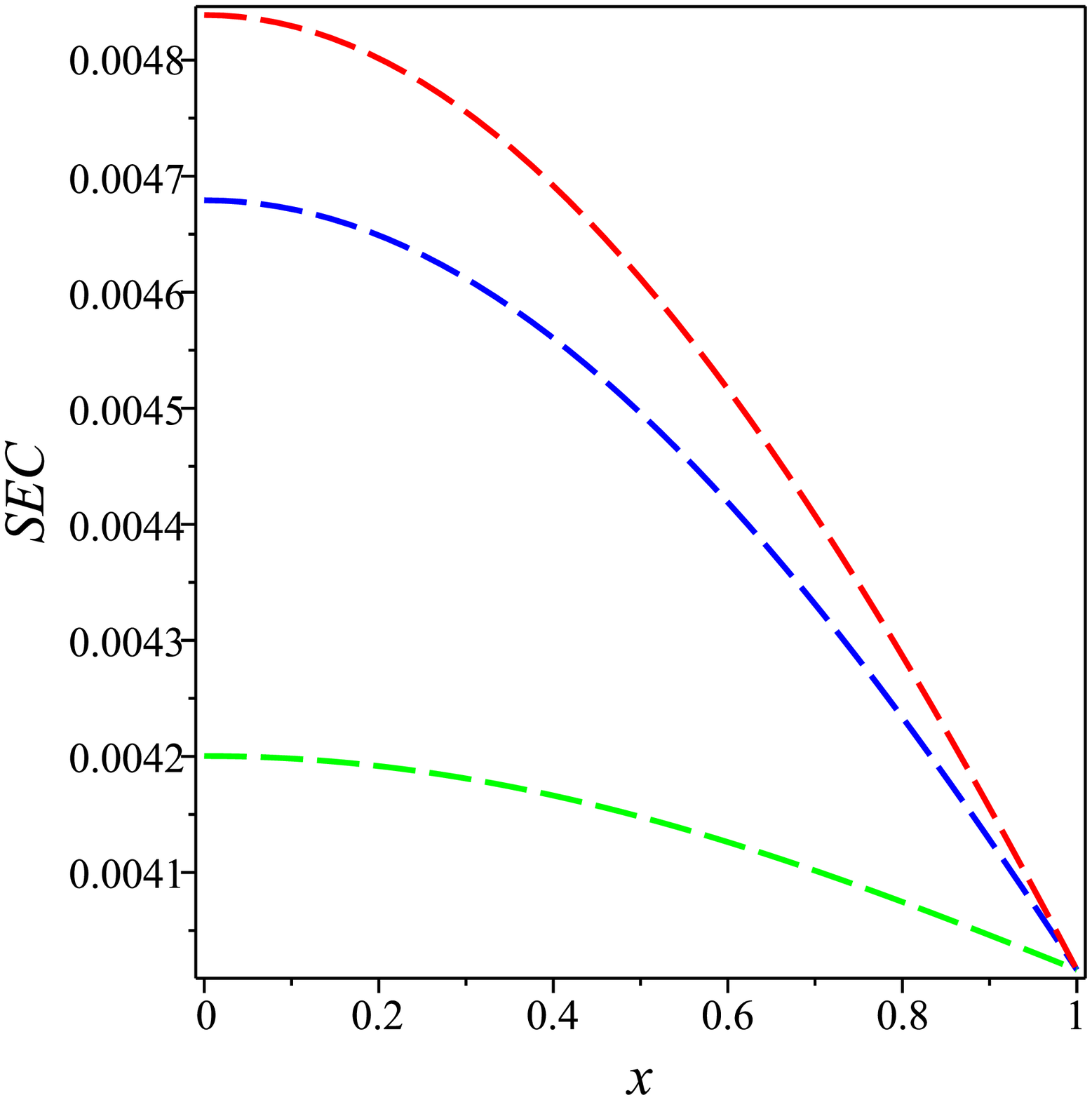}}
\caption[figtopcap]{\small{{{Plots of \subref{fig:grad} gradients of density and pressure, \subref{fig:sped}  speed of sound, \subref{fig:WEC} weak, \subref{fig:SEC0} dominant and \subref{fig:SEC} strong  energy conditions of model (\ref{sol}) , versus the dimensionless x using the constants constrained from \textrm{Cen X-3}.}}}}
\label{Fig:2}
\end{figure}

 Figure \ref{Fig:2} \subref{fig:grad} illustrates that both the gradients of density and pressure are negative. Furthermore, Figure \ref{Fig:2} \subref{fig:sped} demonstrates that the speed of sound is indeed less than unity, which is a necessary condition for a valid stellar model. Additionally, Figures \ref{Fig:2} \subref{fig:WEC}, \subref{fig:SEC0}, and \subref{fig:SEC} exhibit the adherence to energy conditions. Hence, all the criteria associated with energy conditions are fulfilled within the model configuration of Cen X-3, thus meeting the requirements for a true, isotropic, and significant stellar model.
\begin{figure}
\centering
\subfigure[~Equation of state $\omega=\frac{p(x)}{\rho(x)}$ ]{\label{fig:EoS}\includegraphics[scale=0.3]{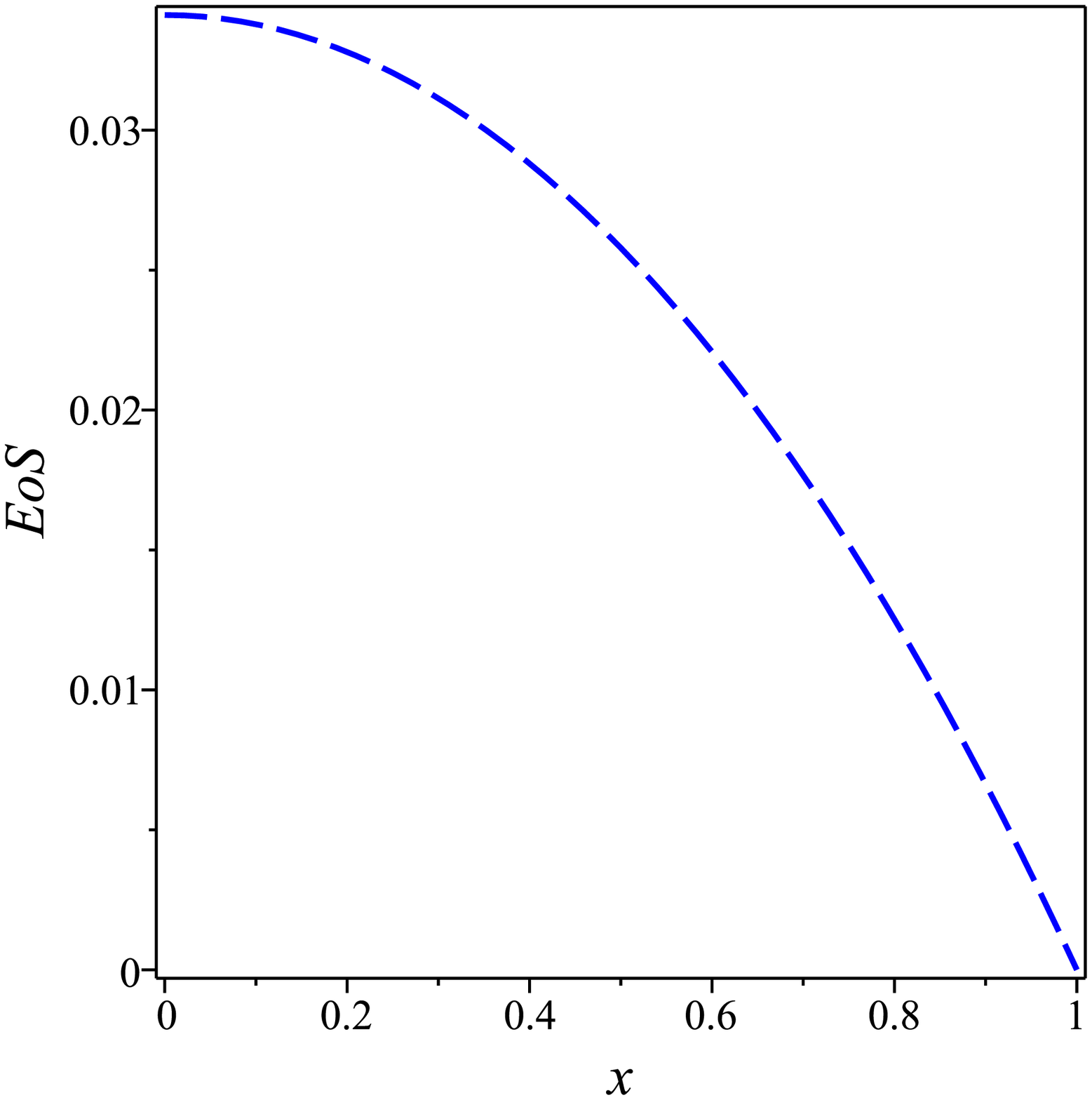}}
\subfigure[~The pressure as function of density i.e., $p(\rho)$]{\label{fig:EoSr}\includegraphics[scale=.3]{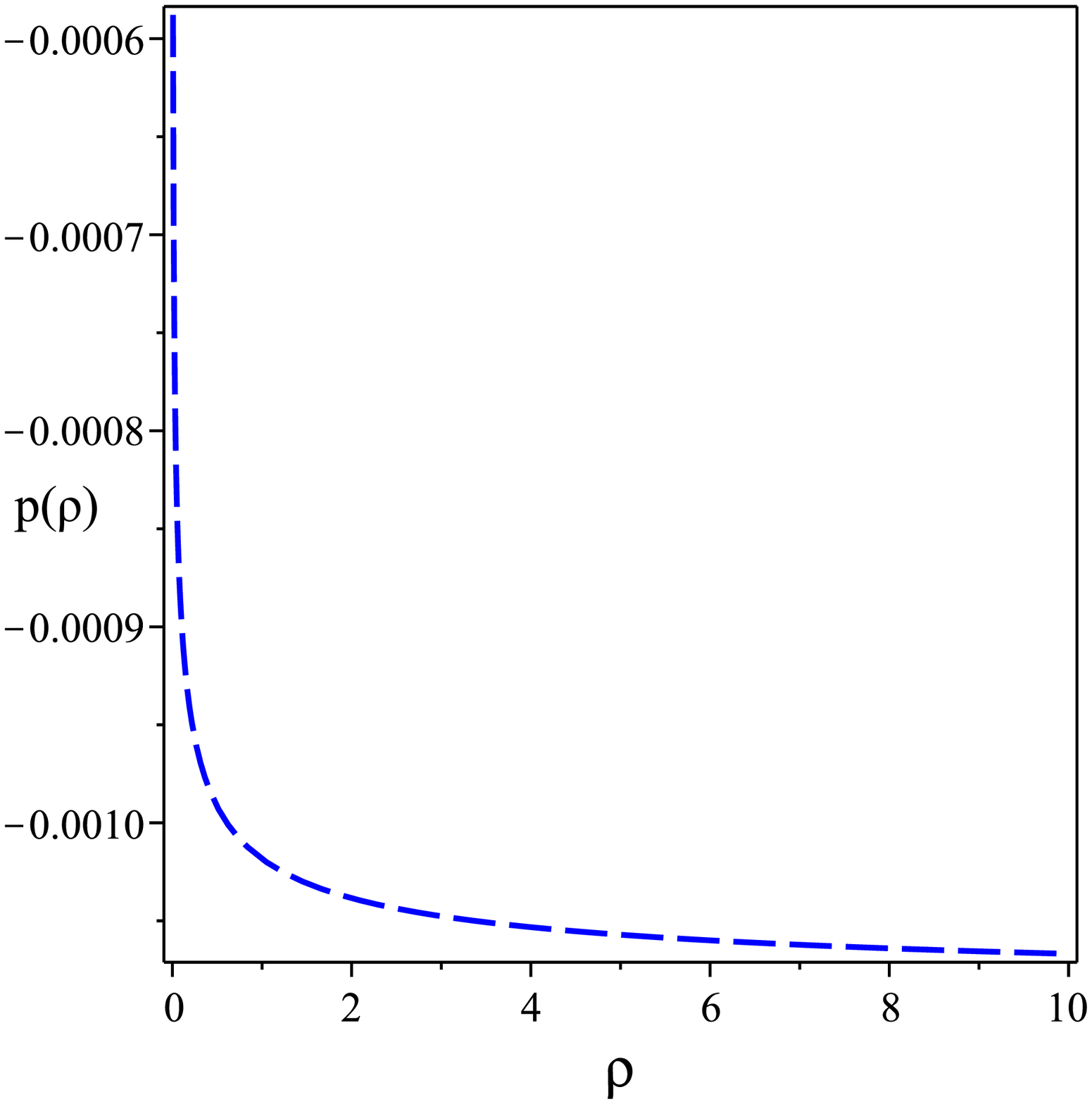}}\\
\subfigure[~The mass and compactness of model (\ref{sol})]{\label{fig:mass}\includegraphics[scale=.3]{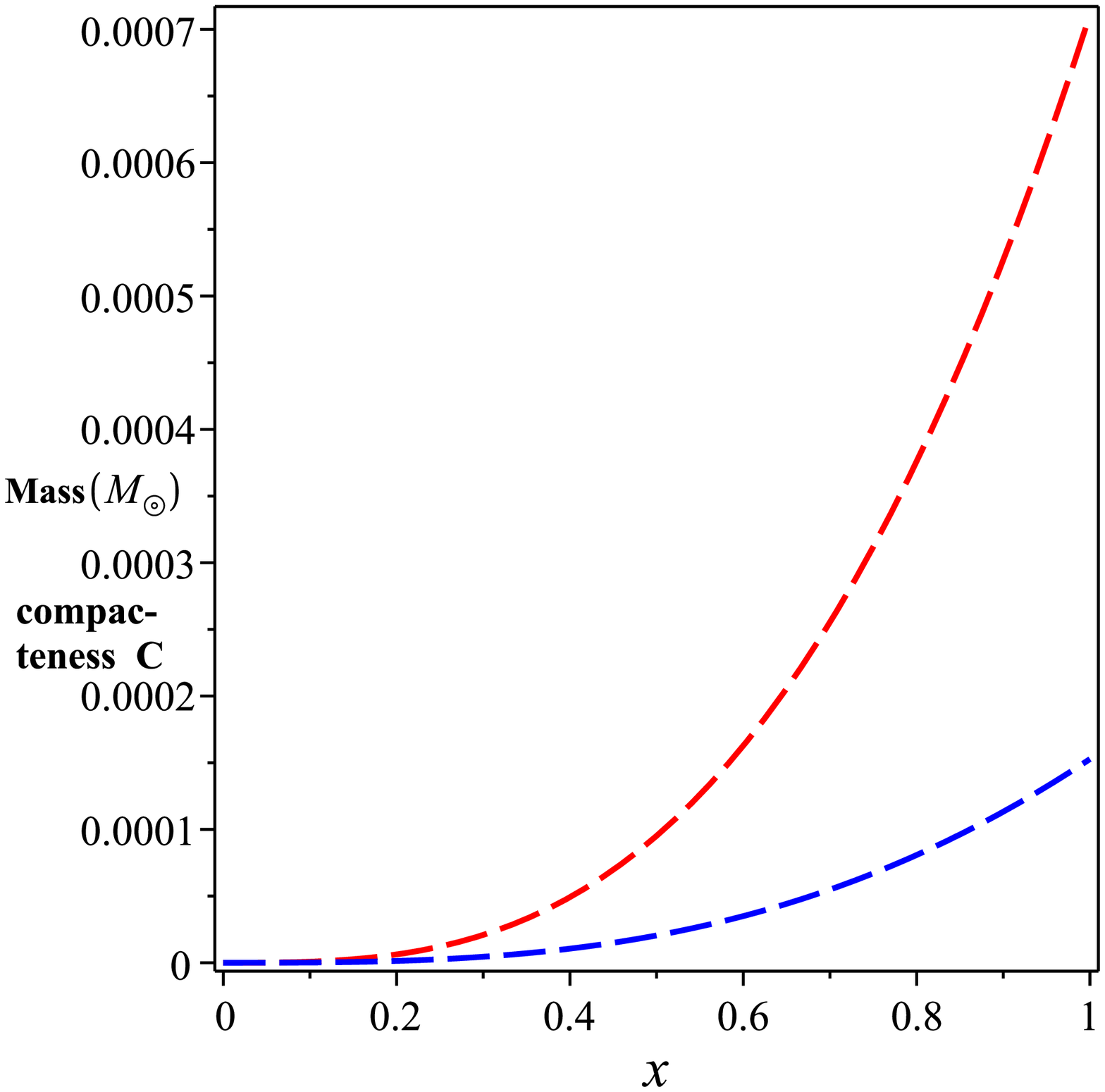}}
\subfigure[~The red shift of the model (\ref{sol})]{\label{fig:red}\includegraphics[scale=.3]{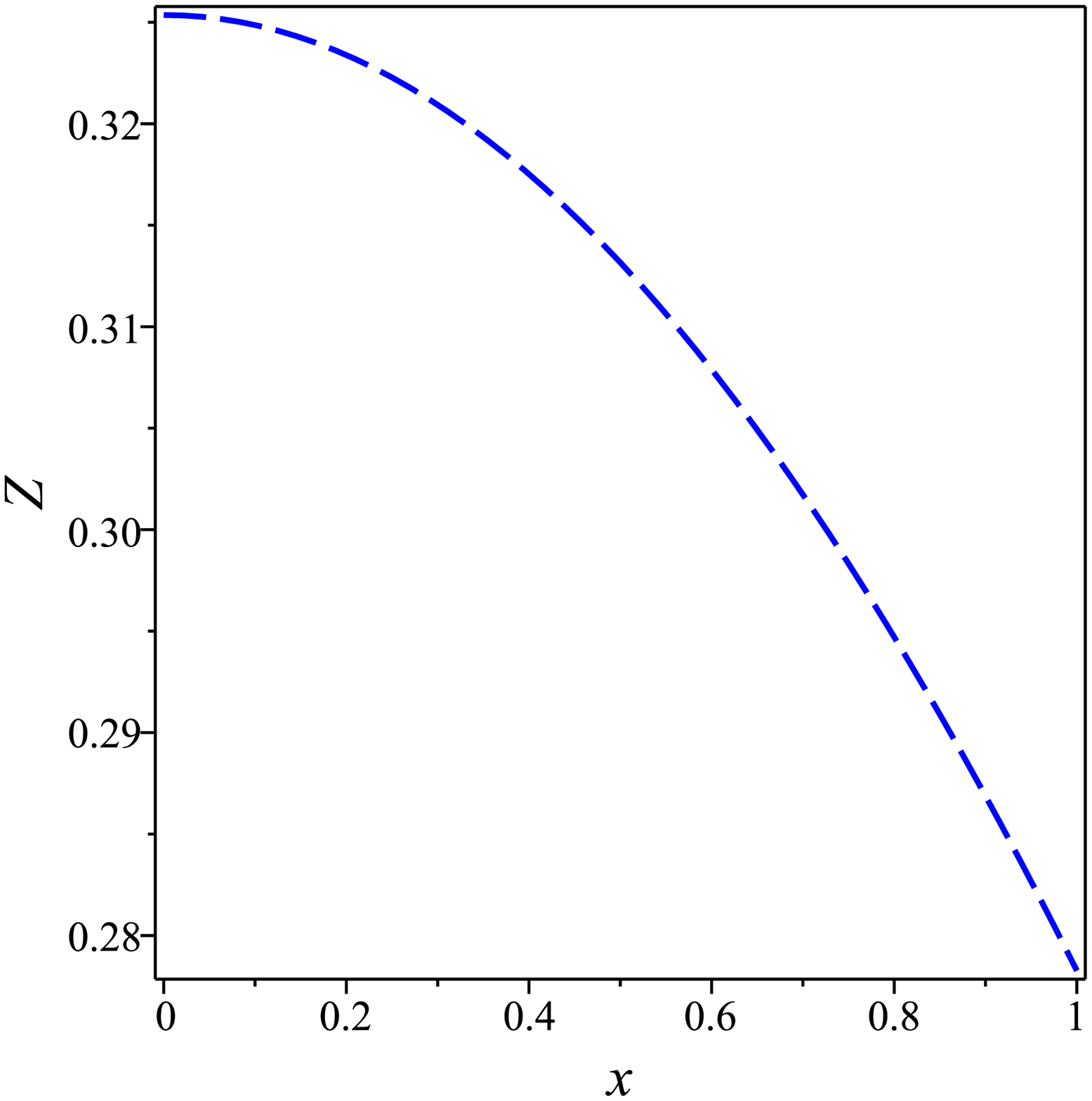}}
\caption[figtopcap]{\small{{Plot of \subref{fig:EoS} the EoS $\omega=\frac{p(x)}{\rho(x)}$ versus the dimensionless x using sing the constants constrained from \textrm{Sen X-3}, \subref{fig:EoSr} the behavior of the pressure as a function of the energy-density,  \subref{fig:mass} the behavior of the mass and compactness and \subref{fig:red} shows the behavior of the red shift.}}}
\label{Fig:3}
\end{figure}

In Fig.  \ref{Fig:3} \subref{fig:EoS} we depict the EoS against the dimensionless $x$ which shows a nonlinear behavior. In Fig.  \ref{Fig:3} \subref{fig:EoSr} we plot the pressure as a function of density which also shows a nonlinear behavior due to the isotropy of model (\ref{sol}). As Figs. \ref{Fig:3} \subref{fig:EoS} and \ref{Fig:3} \subref{fig:EoSr} indicate that the  source  of the non-linearity of the EOS is not the mimetic scalar field  only  but also the isotropy of the stellar model under consideration.

{  The mass function  given  by Eq. (\ref{mas}) is depicted in Fig \ref{Fig:3} \subref{fig:mass}.  Fig. \ref{Fig:3} \subref{fig:mass} show that the behavior of the mass and compactness are  monotonically
increasing of $x$  and $M_{_{x=0}} = 0$.  Moreover, Fig.  \ref{Fig:3} \subref{fig:mass} show   the behavior of the compactness parameter of stellar which are also increasing. Fig. \ref{Fig:3} \subref{fig:mass} shows that the maximum value of the compactness of the \textrm{Cen X-3} is $0.00015$ as shown which is  smaller than the value of GR which is  $0.2035$ \cite{Torres-Sanchez:2019wjv}.}  Finally,  Fig. \ref{Fig:3} \subref{fig:red} indicates the behavior of the red shift of the stellar.
 B\"{o}hmer and Harko \cite{Bohmer2006} limited  the boundary red-shift to be  $Z\leq 5$. The boundary redshift of the model under consideration is evaluated and  get $0.278269891$.
\section{Stability of the model}\label{S7}
We will examine the matter of stability through the utilization of two approaches: the Tolman-Oppenheimer-Volkoff (TOV) equations and the adiabatic index.
\subsection{Equilibrium  using Tolman-Oppenheimer-Volkoff equation}
Now, we  discuss the stability of
the  model (\ref{sol}) by supposing hydrostatic equilibrium
through the TOV  equation.
 TOV equation \cite{PhysRev.55.364,PhysRev.55.374} as  presented  in \cite{PoncedeLeon1993}, gives the following  form of an isotropic model:
\begin{eqnarray}\label{TOV}  -\frac{M_g(x)[\rho(x)+p(x)]{E}}{x\sqrt{E_1}}-\frac{dp}{dx}=0,
 \end{eqnarray}
where $M_g(x)$ is  the gravitational mass which is given by:
\begin{eqnarray}\label{ma}   M_g(x)=4\pi{\int_0}^x\Big({T_t}^t-{T_r}^r-{T_\theta}^\theta-{T_\phi}^\phi\Big)\eta^2E\sqrt{E_1}d\eta=\frac{xE'\sqrt{E_1}}{2E^2}\,,
 \end{eqnarray}
Inserting Eq. (\ref{ma}) into (\ref{TOV}),  we get
\begin{eqnarray}\label{ma1}  -\frac{dp}{dx}-\frac{E'[\rho(x)+p(x)]}{2E}=F_g+F_h=0\,,
 \end{eqnarray}
 with $F_g=-\frac{E'[\rho(x)+p(x)]}{2{E}}$ and $F_h=-\frac{dp(x)}{dx}$ are the gravitational and  the hydrostatic forces respectively.

\begin{figure}
\centering
\subfigure[~TOV equation]{\label{fig:TOVgr}\includegraphics[scale=0.3]{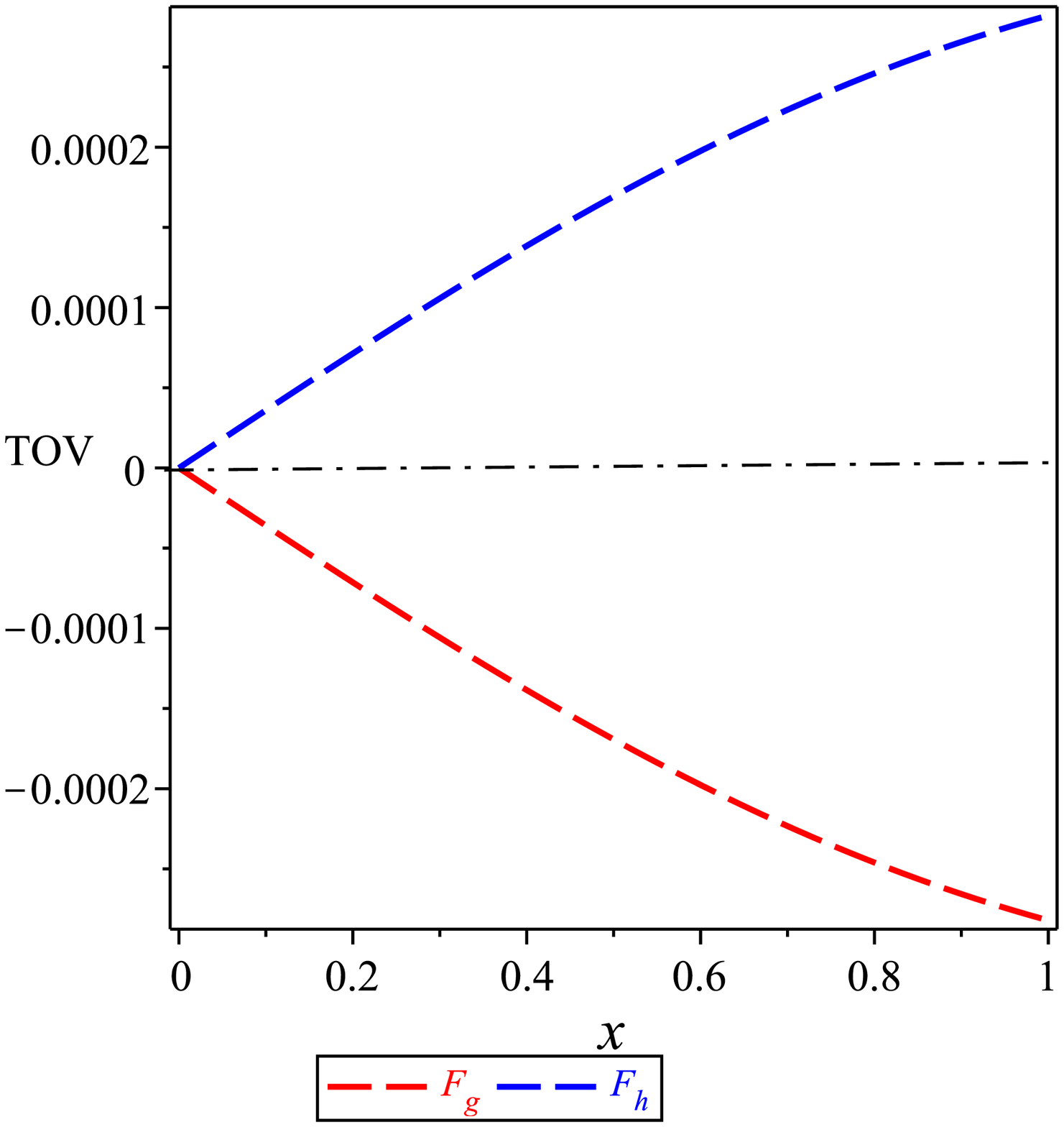}}
\subfigure[~Adiabatic index ]{\label{fig:TOVras}\includegraphics[scale=.3]{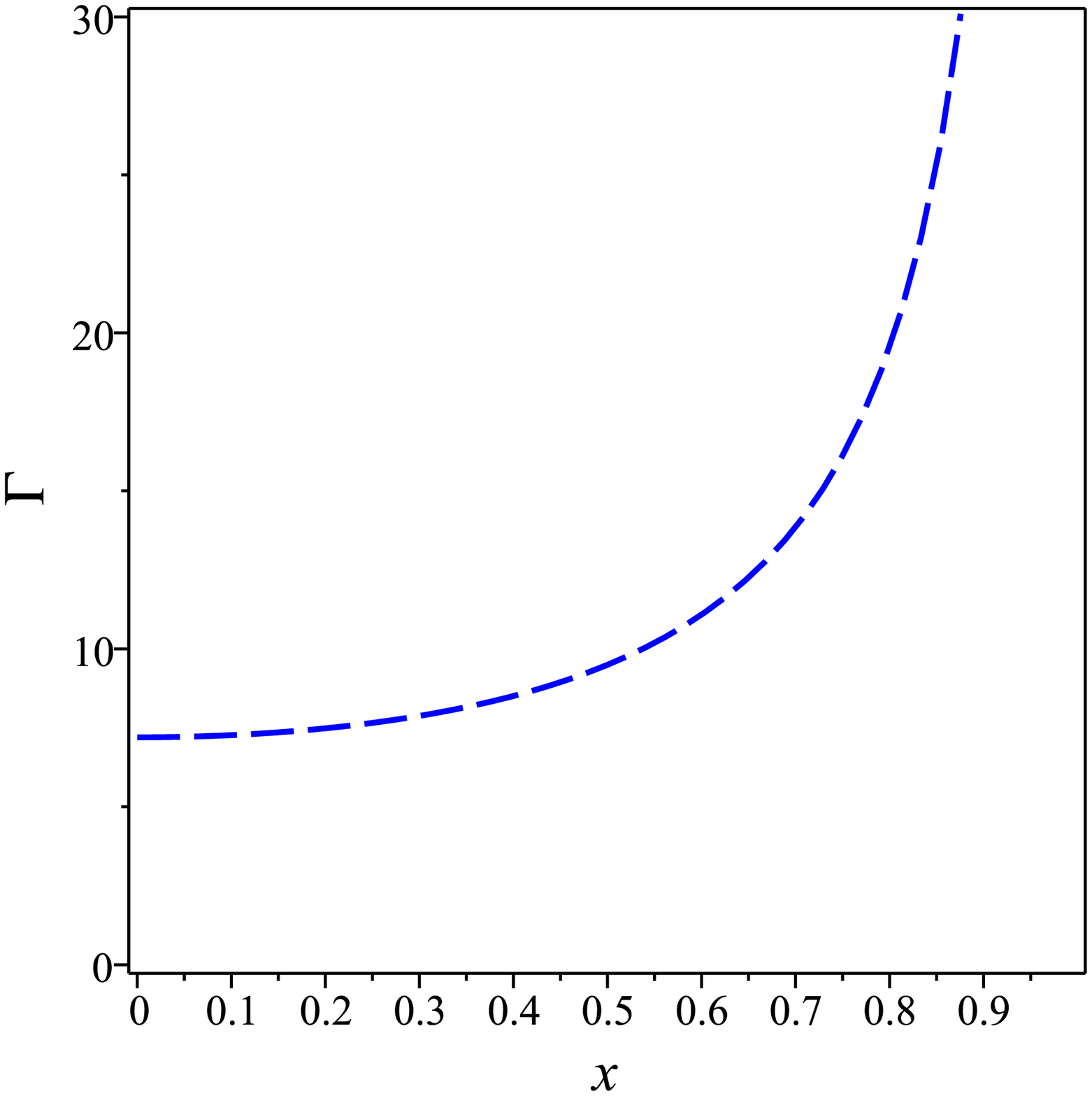}}
\caption[figtopcap]{\small{{Plots of \subref{fig:TOVgr} the TOV equation   \subref{fig:TOVras}  the adiabatic index versus the   dimensionless  $x$.}}}
\label{Fig:4}
\end{figure}

 These two different forces,are plotted in Fig. \ref{Fig:4}.  Therefore,  we prove that the pulsar in static equilibrium is stable through the TOV equation.

\subsection{Adiabatic index}
Another way to examine the stability of the model under consideration  is to study the stability configuration   using the adiabatic index that is considered an essential test. The adiabatic index $\Gamma$ is given by: \cite{1964ApJ...140..417C,1989A&A...221....4M,10.1093/mnras/265.3.533}
\begin{eqnarray}\label{a11}  \Gamma=\left(\frac{\rho(x)+p(x)}{p(x)}\right)\left(\frac{dp(x)}{d\rho(x)}\right)\,.
 \end{eqnarray}
 To have stability
equilibrium the adiabatic index $\Gamma$ must be $\Gamma>\frac{4}{3}$
 \cite{1975A&A....38...51H}. For $\Gamma=\frac{4}{3}$, the isotropic
sphere possesses a neutral equilibrium.
 From Eq. (\ref{a11}),  we obtain the adiabatic index of the model (\ref{sol})  as:
 \begin{eqnarray}\label{aic} &&\Gamma=\frac{3\left( 1+u{x}^{2} \right)  \left( 3+u{x}^{2} \right)\varepsilon}{ \left( 5+u{x}^{2} \right) ^{2}}  \left\{  \left( 12{u}^{2}{x}^{4}+15u{x}^{2}+ 5 \right) (4u\varepsilon+w) \left( 1+u{x}^{2} \right) ^{2}\varepsilon+ \biggl(72 {u}^{3}{x}^{6}+190\,{u}^{2}{x}^{4}+167u{x}^{2}+50 \biggr) \varepsilon^2 u \right\}^{-1}   \nonumber\\
 %
  %
 &&\bigg\{(w-4u\varepsilon_1) \left( 1+u{x}^{2} \right) \left( 24{u}^{3}{x}^{6}+70{u}^{2}{x}^{4}+75u{x}^{2}+50 \right) + \varepsilon u \left(16 {u}^{3}{x}^{ 6}+28{u}^{2}{x}^{4}+40u{x}^{2}+25 \right) \bigg\}^2 \biggl\{3(w-4 u\varepsilon_1) \left( 1+u{x}^{2} \right)\nonumber\\
 && \times\left(16 {u}^{4}{x}^{8}+88{u}^{ 3}{x}^{6}+166{u}^{2}{x}^{4}+115u{x}^{2}+ 25 \right) +u\varepsilon  \biggl( 372{u}^{2}{x}^{4}+32{u}^{4}{x}^{8}+96{u}^{3}{x}^{6} +480u{ x}^{2}+175  \biggr)\biggr\}^{-1} \,.
 \end{eqnarray}


Figure \ref{Fig:4} \subref{fig:TOVras} displays the parameter $\Gamma$, indicating that its values surpass the threshold of $4/3$ within the interior model. This observation confirms that the stability condition is met, as required.
\begin{table*}[t!]
\caption{\label{Table1}%
Values of model parameters}
\begin{ruledtabular}
\begin{tabular*}{\textwidth}{lccccccc}
{{Star}}                 &Reference.            & Mass &       radius [{km}]   &    {$a_0$}    & {$u$} & {$w$}&\\ \hline
&&&&&&&\\
Her X-1     &    \cite{Abubekerov_2008}        &  $0.85\pm 0.15$    &      $8.1\pm0.41$ &     $-0.1738264671$         &$0.1453714459$ &$-2.764746699$     \\
Cen X-3 &\cite{Naik:2011qc}         &  $1.49\pm 0.49$    &  $9.178\pm 0.13$ &$-0.1509026812$&  $0.1241376545$    & $-2.382227207$\\
RX J185635-3754 &\cite{2002ApJ...564..981P}                        &  $0.9\pm 0.2$      &      $\simeq 6$   &  $-0.1583616821$      &$0.1300851651$     &$-2.489973265$ \\
4U1608 - 52  &\cite{1996IAUC.6331....1M}              &  $1.57\pm 0.3$     &  $9.8\pm 1.8$  &     $-0.1637386840$        &$0.1349013201$    &$-2.576874315$  \\
EXO 1745-268  &\cite{Ozel:2008kb}   &  $1.65\pm 0.25$    &  $10.5\pm 1.8$   &     $-0.1653378835$      & $0.1364292317$     &$-2.604379240$ \\
4U 1820-30 &\cite{G_ver_2010}      &  $1.46\pm 0.2$     &  $11.1\pm 1.8$   &    $-0.1713098595$ &$0.1425649058$        &$-2.714525397$\\
&&&&&&&\\

\end{tabular*}
\end{ruledtabular}
\end{table*}
\begin{table*}[t!]
\caption{\label{Table1}%
Values of physical quantities}
\begin{ruledtabular}
\begin{tabular*}{\textwidth}{lcccccccccc}
{{Pulsar}}                              &{$\rho\lvert_{_{_{x\rightarrow0}}}$} &      {$\rho\lvert_{_{_{x\rightarrow 1}}}$} &   {$\frac{dp}{d\rho}\lvert_{_{_{x\rightarrow0}}}$}  &    {$\frac{dp}{d\rho}\lvert_{_{_{x\rightarrow 1}}}$}&
 {$(\rho-3p_r)\lvert_{_{_{x\rightarrow 0}}}$}&{$(\rho-3p)\lvert_{_{_{x\rightarrow 1}}}$}&{$z\lvert_{_{_{x \rightarrow 1}}}$}&\\
  &{[$g/cm^3$]}&     {[$g/cm^3$]} &   {}                &    {}                        &  {[$Pa$]}                &{[$Pa$]}\\
\hline\\[-8pt]
Her X-1                        &$\thickapprox$0.0064     &$\thickapprox$0.0054  &$\thickapprox$0.24 & $\thickapprox$0.25& $\thickapprox$0.0057& $\thickapprox$0.0054& 0.1  \\
Cen X-3           &$\thickapprox$0.0047& $\thickapprox$0.004 &$\thickapprox$0.238 & $\thickapprox$0.246& $\thickapprox$0.0042& $\thickapprox$0.004& 0.28   \\
RX J 1856 -3754             &$\thickapprox$0.011& $\thickapprox$0.01&$\thickapprox$0.238 & $\thickapprox$0.246& $\thickapprox$0.011& $\thickapprox$0.01& 0.22   \\

4U1608 - 52             &$\thickapprox$0.003& $\thickapprox$0.002&$\thickapprox$0.238 & $\thickapprox$0.247& $\thickapprox$0.0029& $\thickapprox$0.0028& 0.18   \\
EXO 1785 - 248   &$\thickapprox$0.0029& $\thickapprox$0.0025&$\thickapprox$0.238 & $\thickapprox$0.247& $\thickapprox$0.0026& $\thickapprox$0.0025& 0.16\\
4U1820 - 30     &$\thickapprox$0.0028& $\thickapprox$0.0023&$\thickapprox$0.238 & $\thickapprox$0.248& $\thickapprox$0.0024& $\thickapprox$0.0023& 0.12      \\

\end{tabular*}
\end{ruledtabular}
\end{table*}

In addition to the pulsar known as$\textrm{ Cen X-3}$, a comparable analysis can be conducted for other pulsars as well. We present concise outcomes for the remaining observed pulsars in Tables I and II.

\section{Discussion and conclusions}\label{S8}
In the present study, we have derived isotropic model of mimetic gravitational theory, for the first time, without assuming any specific form of the EoS. The construction of such model based on the assumption of the metric potential's temporal component and the vanishing of the anisotropy. The main feature of this model was its dependence on three dimensionless constants which we fixed them through the matching condition with the exterior vacuum solution of this theory, i.e., the Schwarzschild solution \cite{Nashed:2018qag}, and the vanishing of the pressure on the surface of the stellar. The physical tests carried out can be summarized: a-The density and pressure must be finite at the center of the stellar configuration, and the pressure must be zero at the surface of the star, Figs. \ref{Fig:1} \subref{fig:dens} and \ref{Fig:1} \subref{fig:pressure}.\\b-The negative values of the gradients of density and pressure Fig.  \ref{Fig:2} \subref{fig:grad}, the validation of the causality Fig. \ref{Fig:2} \subref{fig:sped} as well as its  verification of the energy conditions, Figs. \ref{Fig:2} \subref{fig:WEC}, \subref{fig:SEC0} and \subref{fig:SEC}.\\c-Moreover, we have shown that the EoS parameter, $\omega=\frac{p(x)}{\rho(c)}$ as well as EoS, $p(\rho)=\omega \rho$, behave in a non-linear form which is a feature of the isotropic model, Figs. \ref{Fig:3} \subref{fig:EoS} and   \ref{Fig:3} \subref{fig:EoSr}. Furthermore, we have shown the behavior of the mass and compactness are increasing, and the red-shift of this model has a value on the star's surface as $Z=0.2782$ as shown in  Figs. \ref{Fig:3} \subref{fig:mass} and   \ref{Fig:3} \subref{fig:red}.\\d-One of the merits of this model is that it verified the TOV equation as shown in  Fig. \ref{Fig:4} \subref{fig:TOVgr}  and its behavior  of the adiabatic index is shown in Fig. \ref{Fig:4} \subref{fig:TOVras}.

Additionally, we have examined our model with other six pulsars and derived the numerical values of their constants. Finally, we have derived the numerical values of the density at the center and at the star's surface, the EoS parameter, $\omega$, at the center and the surface of the start, the strong energy condition, and the red-shift at the surface of the stellar configuration. In tables I and II, we tabulated all those data.

To conclude, as far as we know, this is the first time to derive an isotropic model in the framework of the mimetic gravitational theory without assuming any specific form of the EoS. Can this procedure be applied to any other modified gravitational theory like $f(R)$ or $f(T)$? This task will be our coming study.

\section*{Data Availability Statement}
No Data associated in the manuscript.
\section*{Ethics declarations}
Conflict of interest
The author  declares that there is no conflict of interests regarding the publication of this paper.

%

\end{document}